\documentclass{article}
\usepackage{framed,multirow}
\usepackage{amssymb}
\usepackage{amsmath}
\usepackage{latexsym}
\usepackage{url}
\usepackage{hyperref}
\usepackage[margin=0.75in]{geometry}
\usepackage{enumitem}
\usepackage{diagbox}
\usepackage[table, dvipsnames]{xcolor}
\definecolor{titlegray}{gray}{0.75}
\definecolor{lightgray}{gray}{0.75}
\definecolor{lightergray}{gray}{0.95}
\usepackage{soul}
\newcommand{\newtext}[1]{{\color{black}#1}}
\newcommand{\newtextRone}[1]{{\color{black}#1}}
\newcommand{\newtextRtwo}[1]{{\color{black}#1}}
\newcommand{\newtextRtwosecondreview}[1]{{\color{black}#1}}
\newcommand{\newtextRtwothirdreview}[1]{{\color{black}#1}}
\usepackage{graphicx}
\usepackage{wrapfig}
\usepackage{adjustbox}

\DeclareMathOperator*{\argmaximize}{argmax}
\DeclareMathOperator*{\argminimize}{argmin}
\usepackage{chngpage}

\begin{document}

\begin{center}
\Large{Test-time adaptable neural networks for robust medical image segmentation \footnotetext{Published in Medical Image Analysis journal: https://doi.org/10.1016/j.media.2020.101907}}
\vspace{0.5cm}
\\ \large {Neerav Karani, Ertunc Erdil, Krishna Chaitanya, Ender Konukoglu} 
\vspace{0.25cm}
\\ \large {nkarani@vision.ee.ethz.ch} 
\vspace{0.25cm}
\\ \large {Biomedical Image Computing Group, ETH Zurich, Zurich 8092, Switzerland} 
\end{center}

\section*{Abstract}
Convolutional Neural Networks (CNNs) work very well for supervised learning problems when the training dataset is representative of the variations expected to be encountered at test time. In medical image segmentation, this premise is violated when there is a mismatch between training and test images in terms of their acquisition details, such as the scanner model or the protocol. Remarkable performance degradation of CNNs in this scenario is well documented in the literature. To address this problem, we design the segmentation CNN as a concatenation of two sub-networks: a relatively shallow image normalization CNN, followed by a deep CNN that segments the normalized image. We train both these sub-networks using a training dataset, consisting of annotated images from a particular scanner and protocol setting. Now, at test time, we adapt the image normalization sub-network for \emph{each test image}, guided by an implicit prior on the predicted segmentation labels. We employ an independently trained denoising autoencoder (DAE) in order to model such an implicit prior on plausible anatomical segmentation labels. We validate the proposed idea on multi-center Magnetic Resonance imaging datasets of three anatomies: brain, heart and prostate. The proposed test-time adaptation consistently provides performance improvement, demonstrating the promise and generality of the approach. Being agnostic to the architecture of the deep CNN, the second sub-network, the proposed design can be utilized with any segmentation network to increase robustness to variations in imaging scanners and protocols. \newtext{Our code is available at: \url{https://github.com/neerakara/test-time-adaptable-neural-networks-for-domain-generalization}.}

\noindent \textbf{Keywords}: Medical image segmentation, Cross-scanner robustness, Cross-protocol robustness, Domain generalization

\section{Introduction}\label{sec:intro}

\noindent Segmentation of medical images is an important precursor to several clinical analyses. Among the many techniques proposed to automate this tedious task, those based on convolutional neural networks (CNNs) have arguably taken the lead in recent years~\cite{litjens2017survey}.
\newtextRone{Such methods have achieved top performance in several challenges~\cite{litjens2014evaluation, bernard2018deep, zheng2017evaluation}, often outperforming more traditional methods by large margins in accuracy and applicability to multiple problems.}
Indeed, for some anatomies and imaging modalities, the performance of such methods is already comparable to inter-expert variability.
Yet, one of the key issues impeding large-scale adoption of these methods in practice is their lack of robustness to variations in imaging protocols and scanners between training and test images.
\newtextRone{In this work, our goal is to build on the success of segmentation CNNs by increasing their robustness to such changes in their inputs.}

\vspace{0.1cm} \noindent While CNNs are excellent for expressing input-output mappings within the probability distribution corresponding to the training set, they are notorious for responding unpredictably to out-of-distribution inputs - that is, test images that are derived from a different probability distribution~\cite{quionero2009datasetshift}. Such discrepancies in training and test distributions, in other words \emph{domain shifts}, are ubiquitous in medical imaging owing to factors such as changing imaging protocols (e.g. MRI pulse sequences like T1w, T2w, etc.), parameters within the same protocol (e.g. echo time, repetition time, flip angle, etc.), inherent hardware variations in machines manufactured by different vendors, signal-to-noise ratio over time in the same machine. In this context, domains shifts typically manifest as variations in intensity statistics and contrasts between different tissue types. Evidently, CNNs trained for segmentation rely on such low-level intensity characteristics, thereby demonstrating remarkably degraded performance when confronted with such variations at test time~\cite{weese2016four}.

\vspace{0.1cm} \noindent The naive solution of training an independent CNN for each new scanner and protocol is impractical due to the difficulty of repeatedly assembling \textit{large} training datasets.
A somewhat less restrictive scenario is that of \textit{transfer learning}, wherein a CNN, pre-trained on a \textit{source domain} (SD), is fine-tuned with a \textit{few} labelled samples from each new \textit{target domain} (TD)~\cite{van2014transfer, tajbakhsh2016convolutional, karani2018lifelong}.
Further widening the application scope, \textit{unsupervised domain adaptation} \newtextRtwo{(UDA)} relieves the requirement of annotating any TD images at all. Instead, unlabelled TD images are either utilized jointly with the labelled SD dataset directly during the initial training~\cite{kamnitsas2017unsupervised, ouyang2019data, dou2019pnp} or an independent image translation model is learned between the source and target domains~\cite{huo2018synseg}.
Finally, the paradigm of \textit{domain generalization} \newtextRtwo{(DG)} aims to learn a robust input-output mapping using one or more labelled SDs in such a way as to then be \newtextRtwo{also} applicable to unseen TDs~\cite{dou2019domain}.
\newtextRtwo{One of the key distinctions between UDA and DG is that UDA requires the entire SD dataset to be present while training \emph{for each new} TD. We believe that this is a particularly severe requirement in medical imaging, where sharing datasets across institutions often requires regulatory and privacy clearances. DG, on the other hand, requires the SD dataset only for the initial training and not during inference. This is clearly advantageous considering the challenges in data sharing and possibility of encountering a test case that differs from all the domains seen previously. A trained CNN is transported and used to perform inference without requiring access to a labeled or unlabeled training set. As compared to sharing the SD dataset across institutions, it is much easier to transport a trained CNN for usage with images from new TDs.}
We believe that this \newtextRtwo{makes DG} a more practical and promising setting for automating medical image segmentation and therefore, pose our work in this setting.

\vspace{0.1cm} \noindent We hypothesize that in the absence of knowledge about the TD during the initial training, it may be necessary to introduce some adaptability into a segmentation CNN in order to enable it to deal with images arising from new scanners and / or protocols. With this in mind, we propose a segmentation CNN design that concatenates two sub-networks: a relatively shallow image normalization CNN, which we refer to as the \emph{image-to-normalized-image} (I2NI) CNN, followed by a deep CNN that segments the normalized image, which we refer to as the \emph{normalized-image-to-segmentation} (NI2S) CNN. \newtextRtwo{During the training phase, we train both sub-networks jointly, in a supervised fashion, using a SD training dataset. At inference time, we freeze the parameters of NI2S, but adapt those of the I2NI \emph{for each test image}.
During inference, SD samples are not required.} This \emph{test-time adaptation} is driven by requiring that the predicted segmentation be plausible (according to the segmentations observed in the SD dataset), and for dictating such plausibility, we employ denoising autoencoders~\cite{vincent2010stacked} (DAEs).
\newtextRtwo{The test-time adaptation in the proposed method is a part of the inference procedure and does not require any sample other than the test sample at hand. DG through such a test-time adaptation strategy allows adapting a network to any test image independently during inference, which is not possible in UDA and other DG methods.}

\vspace{0.1cm} \noindent To the best of our knowledge, this is the first work in the literature to propose test-time adaptation for tackling the cross-scanner robustness problem in CNN-based medical image segmentation. We believe that the proposed inference time adaptation strategy has the following benefits.
\newtext{Firstly, the proposed method uses a normalization module to adapt to each test image specifically, not relying on the similarity of the test image to previously seen samples as is the case in the majority of the works in the literature.
Secondly, by keeping the adaptable normalization module relatively shallow, we prevent it from introducing substantial structural change in the input image while having sufficient flexibility to correct errors in the predict segmentation.}
Thirdly, as we freeze the majority of the overall parameters at their pre-trained values (those of NI2S), we retain the benefits of the initial supervised training, potentially done with a large number of SD examples and, therefore, valuable for the segmentation task.
Finally, the models used to drive the test-time adaptation, DAEs, can be very expressive as they can potentially exploit high-level cues such as context and shape in order to suggest corrections in the predicted segmentation.
We validate the proposed approach on multi-center MRI datasets from three anatomies (brain, prostate and heart). Experiments demonstrate that the proposed test-time adaptation consistently and substantially improves segmentation performance on unseen TDs over competing methods in the literature.

\section{Related work}
\noindent Improving robustness to scanner and protocol variations in CNN based methods has attracted considerable attention in the literature over the past few years. In the following, we describe some of the dominant approaches proposed either directly in this context, or for the DG problem in other applications.

\vspace{0.1cm} \noindent \textbf{Domain Invariant Features}: 
A dominant approach is to disincentivize reliance on domain-specific signals in the input~\cite{ghifary2015domain, motiian2017unified, li2018domain}. ~\cite{ghifary2015domain} propose to achieve domain invariant features via a separate pre-training step, in which they train a multi-task autoencoder that aims to discover a feature embedding from which all available SDs can be reconstructed. On the other hand,~\cite{motiian2017unified} and~\cite{li2018domain} add regularization losses to enforce feature similarity across domains, with~\cite{motiian2017unified} employing the Maximum Mean Discrepancy~\cite{gretton2007kernel} and~\cite{li2018domain} using an adversarial framework to promote domain invariance. \newtextRtwo{Recently,~\cite{li2018learning,dou2019domain,zhang2020generalizable} proposed a meta-learning framework for encouraging domain independence in the extracted features.}
A common disadvantage of these approaches is the requirement of having access to multiple SDs during training. 

\vspace{0.1cm} \noindent \textbf{Data Augmentation}: 
A related approach is to implicitly encourage domain invariance by expanding the training dataset to include plausible variations that may be encountered at test time~\cite{zhang2019generalizing, jog2019psacnn, volpi2018generalizing, shankar2018generalizing}. This can be achieved by generating simulated input-output pairs by applying heuristic transformations on the available data~\cite{zhang2019generalizing}, by exploiting knowledge about the data generation process~\cite{jog2019psacnn}, by alternately searching for worst-case transformations under the current task model and updating the task model to perform well on data altered with such transformations~\cite{volpi2018generalizing} or by leveraging multiple SDs in order to simulate inputs from \textit{soft, in-between} domains~\cite{shankar2018generalizing}.

\vspace{0.1cm} \noindent \textbf{Imposing Shape Constraints during Training}: 
Another widely suggested idea is to impose shape or topological constraints on the predicted segmentations~\cite{oktay2017anatomically, yue2019cardiac, zheng20183, guo2020improving, chen2019learning, ganaye2018semi, clough2019explicit}. Although not typically proposed for domain generalization, these approaches can be interpreted as imposing invariance in the output space, as opposed to an intermediate feature space. An approach based on this idea has been proposed for unsupervised domain adaptation~\cite{tsai2018learning}. With this viewpoint, it may be possible to train a CNN with multiple SDs and with such regularization on the predicted segmentations.
\newtextRone{These recent works apply the fundamental model-based approach, which is well-established in medical image computing prior to introduction of deep learning methods in~\cite{cootes1995active, pizer2003deformable}, to CNN-based methods}

\vspace{0.1cm} \noindent A common downside of the aforementioned method categories is that they do not offer a way to adapt the CNN at test time. We believe that such adaptability is key in order to deal with unseen target domains and not only rely on the assumption that training set statistics will capture all possible variations that can be encountered at test time. We also note that the methods discussed above are complementary to our approach. They can provide a better base CNN, the NI2S network in our terminology, that our method can further adapt to best suit the given test image.

\vspace{0.1cm} \noindent \textbf{Unsupervised Segmentation}: 
An altogether different approach to circumvent dependence on scanner / protocol specific image characteristics is to not rely on images from a particular SD for learning. \newtextRone{The most successful approaches in this category are based on probabilistic generative models (PGMs)~\cite{van1999automated, zhang2001segmentation, fischl2002whole}.} These methods pose the problem in a Bayesian framework, inferring the posterior probability of the unknown segmentation by specifying a prior model of the underlying tissue classes and a likelihood model, potentially, describing the image formation process. \newtextRone{A downside of these approaches is that they have so far been largely restricted to prior models encoding similarities in relatively small pixel neighbourhoods~\cite{van1999automated, van2001automated, awate2006adaptive, sabuncu2010generative} and mainly used in neuroimaging applications where atlas-based approaches are reliable due to limited morphological variation~\cite{puonti2016fast, dalca2019unsupervised}}. Recent works leverage a set of segmentations in order to learn long-range spatial regularization priors through Markov random fields with high-order clique potentials~\cite{agn2019modality, brudfors2019nonlinear} as well as through variational auto-encoders~\cite{dalca2018anatomical}. Nevertheless, most of these methods involve deformable image registration as one of their pre-processing steps, thus making it challenging to extend them to applications beyond neuroimaging.

\vspace{0.1cm} \noindent \textbf{Test Time Adaptation}: 
Adaptation of a pre-trained CNN for each test image has been proposed recently~\cite{wang2018interactive, zhang2020fidelity, sun2019test} for different applications. \cite{wang2018interactive} suggest such adaptation for interactive segmentation of unseen objects and drive the adaptation using a smoothness based prior on the predicted segmentation. In the context of undersampled MRI reconstruction,~\cite{zhang2020fidelity} suggest fine-tuning a pre-trained CNN, such that the predicted image reconstruction is consistent with a known forward image reconstruction model. Finally,~\cite{sun2019test} tackle the domain generalization problem and propose to adapt a part of their task CNN according to a self-supervised loss defined on the given test image.

\vspace{0.1cm} \noindent \textbf{Post-Processing}:
Finally, it has been suggested to post-process CNN predictions, according to a smoothness prior defined using conditional random fields~\cite{krahenbuhl2011efficient, kamnitsas2017efficient} or a prior based on denoising autoencoders~\cite{larrazabal2019anatomical} or generative adversarial networks~\cite{engin2020agan}.
While such post-processing steps provide a way to improve plausibility of the predicted segmentations, they might lead to a situation wherein the post-processed segmentation is inaccurate for the given input image.

\section{Method}\label{sec:method}

\noindent Let $X$ and $Z$ be random variables denoting images and segmentations / labels (we use these two terms interchangeably), respectively. We assume that we have access to a dataset $\mathcal{D}_{SD}$: $\{(x_{i}, z_{i}^{}) | \ \ i = 1, 2,\dots N \}$, where $x_{i} \sim P_{SD}(X)$ are samples from a SD and $z_{i}$ are corresponding ground truth segmentations.
The $\mathcal{D}_{SD}$ can be composed of images coming from only one scanner and protocol setting or contain images from multiple scanners and protocols. The proposed method is agnostic to the formation of $\mathcal{D}_{SD}$.
In our experiments, in view of potential difficulties in annotating and aggregating data from multiple imaging centers, we restrict the SD to a particular combination of imaging scanner and protocol setting. 
Given this annotated dataset, the goal is to provide an automatic segmentation method that works for new images sampled from not only the SD ($P_{SD}(X)$), but also unseen TDs ($P_{TD}(X)$).

\subsection{Training a Segmentation CNN on the Source Domain}~\label{sec:initial_training}
\noindent Firstly, we train a segmentation CNN, segCNN, using the SD training dataset, $\mathcal{D}_{SD}$. Formally, we use two CNNs to model the transformation from the image space to the space of segmentations, $Z = S_{\theta}(N_{\phi} (X))$. $N_{\phi}$ and $S_{\theta}$ are I2NI and NI2S mentioned in the introduction, respectively, and segCNN is defined as their concatenation.
The optimal parameters of segCNN, $\{\theta^*, \phi^*\}$, are estimated by minimizing a supervised loss function:

\begin{equation}\label{eqn_dis_seg__estimate_params}
    \theta^*, \phi^*
    = \argminimize_{\theta, \phi} \sum_{i}^{} L(S_{\theta}(N_{\phi} (x_i)) , z_i)
\end{equation}

\noindent where $\{x_i, z_i^{}\}$ are image-label pairs from $\mathcal{D}_{SD}$,
the sum is over all such pairs used for training and $L$ is a loss function that measures dissimilarity between the ground truth labels and predictions of the network. \newtext{In non-adaptable networks, the common application of CNNs for segmentation, once the optimal parameters of segCNN are estimated, the segmentation for a new image $x$ is obtained as $z^{*} = S_{\theta^*}(N_{\phi^*}(x)) $. In this work, we modify this procedure by introducing an adaptation step at test time.}

\subsection{Test-Time Adaptation}~\label{sec:test_time_adaptation}
\noindent At inference time, when confronted with shifts in the input distribution, the mapping described by the pre-trained segCNN may not be reliable because the optimization in Eq.~\ref{eqn_dis_seg__estimate_params} depends on the SD training dataset, and in particular, on the intensity statistics of the SD images, $x_{i} \sim P_{SD}(X)$. To address this, we propose to use the pre-trained values of the segCNN parameters as an initial estimate, further adapting them for each test image. In order to implement this idea, we have to make two design choices: (1) \newtextRtwo{which parameters to update at test-time} and (2) how to drive such an update, without label information and with only the test image available.

\vspace{0.1cm}
\subsubsection{\newtextRtwo{Which parameters of segCNN to adapt at test-time?}}~\label{subsec:test_time_adaptation_only_normalization_module}
\noindent To answer this question, we assume that domain shifts due to changing imaging protocols and scanners manifest in the form of differences in low-level intensity statistics and contrast changes between different tissue types. Accordingly, we posit that a relatively shallow image-specific normalization module might provide sufficient adaptability to obtain accurate segmentations within the relevant domain shifts. This reasoning underlies the formulation of segCNN as a concatenation of two transformations: $Z = S_{\theta}(X_{n})$ and $X_{n} = N_{\phi}(X)$, $X_n$ being normalized images. Here, $N_{\phi}$ denotes a normalization (I2NI) CNN, with parameters $\phi$ that are initialized with pre-trained values $\phi^*$ and further adapted for each test image, while $S_{\theta}$ denotes a normalized-image-to-segmentation (NI2S) CNN, with parameters $\theta$ that are fixed at their pre-trained values $\theta^*$ at test time.

\vspace{0.1cm} \noindent We model $N_{\phi}$ as a residual CNN. It processes the input image with $n_N$ convolutional layers, each with kernel size $k_N$ and stride 1. We employ no spatial down-sampling or up-sampling in $N_{\phi}$ and have it output the same number of channels as the input image. We hypothesize that such an adaptable normalization module could enable an image-specific intensity transformation in order to alter the TD image's contrast such that the pre-trained NI2S CNN, $S_{\theta^{*}}$, can accurately carry out the segmentation. 

\vspace{0.1cm} \noindent \newtextRone{Simultaneously, by restricting the kernel size ($k_N$) as well as the number of layers ($n_N$) to relatively small values, we limit $N_{\phi}$ to expressing intensity transformations that are sufficient for modeling contrast changes, but insufficient for substantially altering the image content by adding, removing or moving anatomical structures.}
Further, we believe that an important benefit of our formulation is that it freezes the majority of the overall parameters (those of the deep NI2S CNN) at their pre-trained values, thus essentially leveraging the NI2S segmentation network at its full capacity using the weights determined through supervised learning, as described in Sec.~\ref{sec:initial_training}.


\vspace{0.1cm} \subsubsection{How to drive the test-time adaptation?}~\label{subsec:test_time_adaptation_how_to_drive}
\newtext{The main challenge in test-time adaptation is the lack of label information and additional images. The model only has access to the test image to which it should adapt.}
In our method, the main aim that drives the adaptation is to produce segmentations that are plausible, that is, similar to those seen in the SD training dataset. 
The underlying assumption here is that the domain shifts in question pertain only to scanner and protocol changes, with the images otherwise containing similar structures, whether healthy or abnormal, as the SD dataset. 
To this end, we use denoising autoencoders (DAEs)~\cite{vincent2010stacked} to assess the similarity of a given segmentation to those in the SD dataset.
The idea is that if the segmentation predicted by segCNN is implausible, the DAE will see it as a "noisy" segmentation and "denoise" it to produce a corresponding plausible segmentation. The output of the DAE can then be used to drive the aforementioned test-time adaptation.
Crucially, DAEs can be highly expressive - they have the capacity to leverage high-level cues, such as long-range spatial context and shape, in order to suggest corrections in segCNN's predictions.

\begin{figure}[h!]
\centering
    \includegraphics[trim = 0mm 0mm 0mm 0mm, angle=0, clip, width=0.48\textwidth]{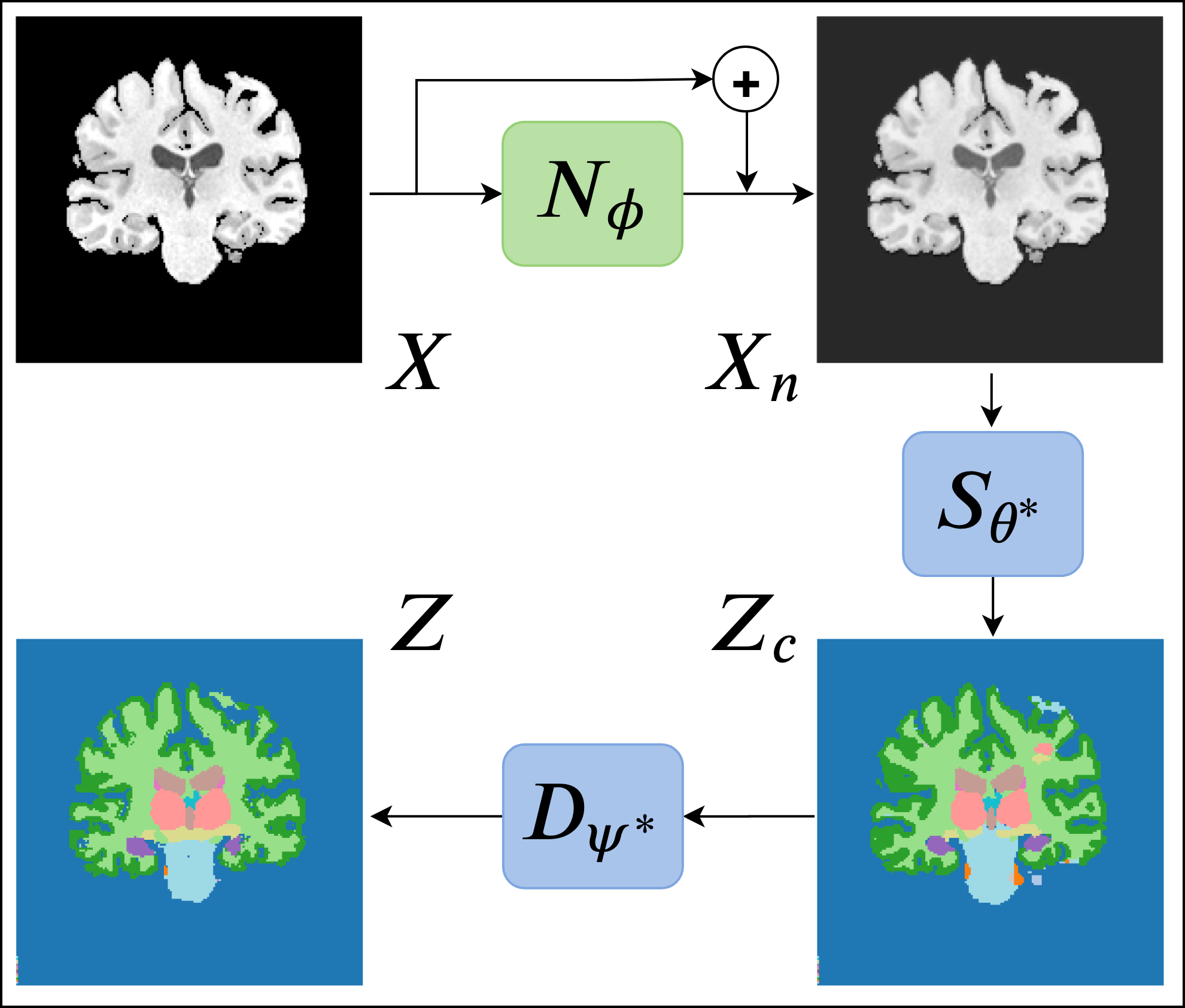}
\caption{Workflow of the method: For each test image, $N_\phi$ is adapted such that the resulting segmentation is plausible, as gauged by a DAE, $D_\psi^*$. The modules shown in blue are trained on the SD and fixed thereafter, while the green module is adapted for each test image.}\label{fig_workflow}
\end{figure}

\vspace{0.1cm} \noindent The workflow of our test-time adaptation method is depicted in Fig.~\ref{fig_workflow}.
We leverage the available ground truth segmentations in the SD training dataset, $\mathcal{D}_{SD}$, to train a DAE, $D_{\psi^*}$, that maps corrupted segmentations $Z_c$, which are not necessarily similar to those in the SD training dataset, to ``denoised'' segmentations $Z$, similar to those in the SD training dataset. The details of this training are explained in Sec.~\ref{sec:dae_training}. For the time being, let us assume that we have a trained DAE, $D_{\psi^*}$. 
For a given test image $x$ and a set of parameters for the I2NI CNN, $\phi$, we treat the segmentation predicted by $S_{\theta^*}(N_{\phi}(x))$ as a ``noisy'' or ``corrupted'' segmentation.
We pass this noisy segmentation through $D_{\psi^*}$ and obtain its denoised version. Now, we update the parameters of the adaptable I2NI CNN, $N_{\phi}$, so as to pull the predicted segmentation closer to its denoised version:
\begin{equation}\label{eqn_tta}
    \hat{\phi}
    = \argminimize_{\phi} 
    L(z_c, D_{\psi^*}(z_c));\ z_c = S_{\theta^*}(N_{\phi}(x)),
\end{equation}

\noindent 
where $L$ is a similar loss to that in Equation~\ref{eqn_dis_seg__estimate_params}. 
Eqn.~\ref{eqn_tta} denotes the test-time adaptation that we carry out for each test image $x$. This optimization is done iteratively (using either gradient descent or a variant thereof). At the beginning of the optimization for a test image, segCNN trained on the SD likely predicts a corrupted segmentation if the test image is not from SD, such as the one shown on the bottom-right in Fig.~\ref{fig_workflow}. The DAE takes this prediction as input and proposes a corrected segmentation, such as the one shown on the bottom-left in Fig.~\ref{fig_workflow}. Now, the parameters of the adaptable I2NI CNN, $N_\phi$, are updated so as to minimize the dissimilarity between the DAE input and output. As the optimization proceeds, the segmentation predicted by segCNN becomes increasingly plausible, that is, similar to those in SD training dataset. Therefore, the DAE input and output become similar, resulting in small loss values and convergence of the test-time adaptation.
\newtextRone{Importantly, the adaptable normalization module $N_\phi$ is relatively shallow and has a relatively small receptive field. Thus, the adaptation cannot introduce large structure alterations but it is free to change the contrast of the input image.
The optimization runs for a pre-specified number of iterations and the optimal I2NI parameters $\hat{\phi}$ are chosen as the ones that provide the least dissimilarity between the DAE input $z_c$ and output $D_{\psi^*}(z_c)$} during the iterations. The final segmentation is predicted as $\hat{z} = S_{\theta^*}(N_{\hat{\phi}}(x))$.

\subsection{DAE Training}~\label{sec:dae_training}
\noindent The DAE described above is a key component for driving the test-time adaptation. 
We model it as a 3D CNN - this potentially allows for learning of information about relative locations of different anatomical structures across volumetric segmentations as well as about their shapes in their entirety. In order to train such a DAE, we generate a training dataset of pairs ($z_{i}, z_{ci}$), with $z_{i} \sim P(Z)$ (available from $\mathcal{D}_{SD}$) and $z_{ci} \sim P(Z_{c} | Z = z_i; \omega)$, a corruption process that we define in order to generate corrupted segmentations $Z_c$ given clean segmentations $Z$. With this dataset, we train the DAE to predict $Z = D_{\psi}(Z_{c})$ by minimizing the following loss function to estimate the parameters $\psi^*$:
\begin{equation}\label{eqn_dae_training1}
    \psi^*
    = \argmaximize_{\psi} \mathbb{E} [L(D_{\psi}(Z_{c}) , Z)]
\end{equation}
\noindent Here, the expectation is over the joint distribution $P(Z, Z_{c}) = P(Z)P(Z_{c} | Z)$. Thus, we have
\begin{equation}\label{eqn_dae_training2}
    \psi^*
    = \argminimize_{\psi} \sum_{j} \sum_{i} L(D_{\psi}(z_{cij}) , z_i)
\end{equation}
\noindent where the index $j$ denotes different samples obtained from $P(Z_{c} | Z = z_i; \omega)$, the outer sum is over the number of corrupted samples that we generate for each ground truth label $z_i$ and $L$ is a loss function that computes dissimilarity between the clean ground truth labels and the predictions of the DAE.

\vspace{0.1cm} \noindent \textbf{Noising Strategy}: The main design choice for the DAE training described above is the noising process, $P(Z_{c} | Z; \omega)$. This noising process is used to generate artificially degraded segmentations, simulating the inaccurate labels that pre-trained segCNN will likely predict when faced with input images from unseen TDs.
In this work, we follow a heuristic procedure for generating such noisy labels: we copy cubic patches from randomly chosen locations in the label image to other randomly chosen locations in the same image.
\footnote{If the noising process is chosen to be one that adds Gaussian noise to its inputs and if the DAE is trained by minimizing the $L_2$ loss (i.e if $L(D_{\psi}(z_{cij}) , z_i) = || D_{\psi}(z_{cij}) - z_i ||_2$), then the gradient of the label prior, $P(Z)$, can be expressed in terms of the DAE reconstruction error~\cite{bigdeli2017deep}. This allows for explicit prior maximization~\cite{wang2019denoising}. However, this result does not generalize to different data corruption models, such as the noising strategy used in this work. On the other hand, a simple noising model that adds Gaussian noise is unlikely to mimic the inaccurate segmentations predicted for TD images by the pre-trained segCNN.}
In each training iteration of the DAE and for each clean label, the number of such patches ($n_1$) is sampled from an uniform distribution $U(0, n_{1}^{max})$. For each of these $n_1$ patches, its size ($n_2$) is sampled independently from another uniform distributionn $U(0, n_{2}^{max})$. Thus, our noising process is defined by hyper-parameters: $\omega: \{n_{1}^{max}, n_{2}^{max}\}$.

\subsection{Atlas initialization for test-time adaptation for large domain shifts}~\label{sec:atlas_initial_training}
\noindent We note that the DAE could itself be vulnerable to domain shifts in its inputs. Such a risk can be mitigated if the probability distribution of the corrupted segmentations generated by our noising process approximates that of the predictions of the pre-trained segCNN in response to TD images. For domain shifts pertaining to scanner changes under the same imaging protocol, we assume that our noising process is able to satisfy this requirement.
Specifically, for a given test image $x$, the segmentation predicted by the pre-trained segCNN as well as during the iterative test-time adaptation is $z_{c} = S_{\theta^*}(N_{\phi}(x))$. Now, if $x$ is acquired using the same imaging modality and similar protocol as the SD images and has unknown ground truth segmentation $z$, then we assume that $z_{c}$ can be seen as a corrupted segmentation that is a sample from our noising process $P(Z_c | Z = z)$.

\vspace{0.1cm} \noindent On the other hand, this assumption is violated when SD and TD images are very different, for instance when acquired using different modalities or very different protocols, such as using MR for one image and CT for the other or T1-weighted MR for one and T2-weighted for the other. 
In such cases, the predictions of the pre-trained segCNN can be highly corrupted and may not be captured by the noising strategy described above, and thus the corresponding DAE outputs may no longer be reliable for driving the test-time adaptation. Therefore, when dealing with large domain shifts consisting of imaging protocol changes, we utilize an affinely registered atlas, $A$, to first draw the predicted segmentations to a reasonable starting point from where the DAE can take over. Specifically, instead of directly carrying out the optimization as described in Eq.~\ref{eqn_tta}, we switch between minimizing $L(z_c, D_{\psi^*}(z_c))$ and $L(z_c, A)$, both with respect to $\phi$. Here, $z_c = S_{\theta^*}(N_{\phi}(x))$ are the predictions of segCNN at any point during the iterative test-time adaptation. For deciding when to make this switch, we employ a threshold-based approach: if $d(z_c, D_{\psi^*}(z_c)) / d(z_c, A) \geq \alpha$ and $d(z_c, A) \geq \beta$, then we minimize $L(z_c, D_{\psi^*}(z_c))$. Else, we minimize $L(z_c, A)$. Here, $d$ is a similarity measure between segmentations and $\alpha, \beta$ are hyper-parameters. In our experiments, we use the Dice loss~\cite{milletari2016v} as $L$ and the Dice score as $d$, with $L = 1-d$. The reasoning for the threshold-based switching is as follows: In the initial steps of the test-time adaptation, the predicted segmentations will likely be extremely corrupted. Therefore, we would like to use the affinely registered atlas for driving the adaptation. Once the predicted segmentations improve and can be considered as samples from our noising process, we would like to switch to using the DAE outputs for driving the adaptation, as it has more flexibility than an affinely registered atlas. Our threshold-based switching procedure encodes two signals indicating improvement in the predicted segmentations: increased similarity between (1) DAE input and output (note that when the predicted segmentation is plausible according to the DAE, the DAE models an identity transformation) and (2) predicted segmentation and the atlas.


\subsection{Integrating 2D Segmentation CNN with 3D DAE}~\label{sec:gradient_accumulation}
\noindent As noted in Sec.~\ref{sec:dae_training}, we model the DAE as a 3D CNN in order to leverage 3D information. On the other hand, the CNN-based image segmentation literature is dominated by 2D CNN designs, mainly because 3D CNNs are hindered by memory issues and 2D CNNs already provide state-of-the-art segmentation performance in many cases~\cite{baumgartner2017exploration}. In order for our test-time adaptation method to be applicable to both 3D as well as 2D segmentation CNNs, we propose the following strategy for using 2D segmentation CNNs with our 3D DAE. In this case, the normalization $N_{\phi}$ network is also a 2D CNN and we iteratively carry out the following two steps for T updates of the parameters of $N_\phi$ for a given 3D test image:
\begin{enumerate}
    \item Predict the current segmentation for the entire 3D test image, by passing it through the 2D segmentation CNN in batches consisting of successive slices. Following this, pass the 3D predicted segmentation though the trained DAE to obtain its denoised version.
    \item Initialize gradients with respect to $\phi$ to zero. Process the 3D test image in 2D batches consisting of successive slices as in step 1: For each batch, predict its segmentation, compute loss between the prediction and the corresponding batch of the denoised labels computed in step 1, and maintain a running sum of gradients of the loss with respect to $\phi$. At the end of all batches, average gradients over the number of batches and update $\phi$.
\end{enumerate}

\noindent If the domain shift includes a change in imaging protocol, we use the threshold based method described in Sec.~\ref{sec:atlas_initial_training} to determine whether to use the atlas or the DAE outputs as target labels for driving the adaptation in step 1. Furthermore, to save computation time, we update the denoised labels for the adaptation, i.e. run step 1, after $f$ runs of step 2, instead of after every run.
\section{Experiments and Results}\label{sec:exp_and_results}

\subsection{Datasets}\label{sec:datasets}
\noindent We validate the proposed method on three anatomies, the brain, heart and prostate, and multiple MRI datasets. \newtextRtwo{Noting that aggregating datasets across multiple imaging centers might be difficult in practice, in our experiments we consider the domain generalization problem under the constraint of availability of a labelled dataset from only one SD, potentially consisting of images from multiple scanners.}

\vspace{0.1cm} \noindent \textbf{Brain MRI}: For brain segmentation, we use images from 2 publicly available datasets: Human Connectome Project (HCP)~\cite{van2013wu} and Autism Brain Imaging Data Exchange (ABIDE)~\cite{di2014autism}. In the HCP dataset, both T1w and T2w images are available for each subject, while the ABIDE dataset consists of T1w images from several imaging sites. We use the HCP-T1w dataset as the SD and the ABIDE-Caltech-T1w and HCP-T2w datasets as TDs. In the experiments with brain MRI, we segment the following 15 labels: background, cerebellum gray matter, cerebellum white matter, cerebral gray matter, cerebral white matter, thalamus, hippocampus, amygdala, ventricles, caudate, putamen, pallidum, ventral DC, CSF and brain stem.
    
\vspace{0.1cm} \noindent While providing great imaging data in large quantities, unfortunately, these datasets do not provide manual segmentation labels. Moreover, we are not aware of publicly available brain MRI datasets in large quantities with manual segmentations of multiple subcortical structures. In this situation, we employ the widely used FreeSurfer~\cite{fischl2012freesurfer} tool to generate pseudo ground truth segmentations for the HCP and ABIDE datasets. FreeSurfer is an example of a successful segmentation tool, that works robustly across scanner and protocol variations. However, it has the downside of being excessively time expensive, taking as much as 10 hours on a CPU for segmenting one 3D MR image, and specific to the brain.

\vspace{0.1cm} \noindent \textbf{Prostate MRI}: For prostate segmentation, we use the National Cancer Institute (NCI) dataset~\cite{nci_prostate_dataset} as the SD. As TDs, we use the PROMISE12 dataset~\cite{litjens2014evaluation} and a private dataset obtained from the University Hospital of Zurich (USZ)~\cite{Becker2019a}. For the NCI and USZ datasets, expert annotations are available for 3 labels for each image: background, central gland (CG) and peripheral zone (PZ), while the PROMISE12 dataset only provides expert annotations for the whole prostate gland (CG + PZ). Thus, we evaluate our predictions both for the whole gland as well as separate CG and PZ segmentations, respectively, for the different datasets.
    
    
\vspace{0.1cm} \noindent \textbf{Cardiac MRI}: For cardiac segmentation, we use the Automated Cardiac Diagnosis Challenge (ACDC) dataset~\cite{bernard2018deep} as SD and the right ventricle segmentation challenge (RVSC) dataset~\cite{petitjean2015right} as TD. Annotations are available for LV (left ventricle) endocardium, LV epicardium and RV (right ventricle) endocardium for the ACDC dataset, and for the RV epicardium and RV endocardium for the RVSC dataset. Thus, the segmentation CNN, trained on the ACDC dataset and adapted for the RVSC dataset, segments LV endocardium, LV epicardium and RV endocardium. However, we evaluate the domain generalization performance based on RV endocardium segmentation, the only structure that is common in both datasets, setting other predictions as background.
    
\begin{figure}[h!]
\centering
    \includegraphics[trim = 0mm 0mm 0mm 0mm, angle=0, clip, width=0.48\textwidth]{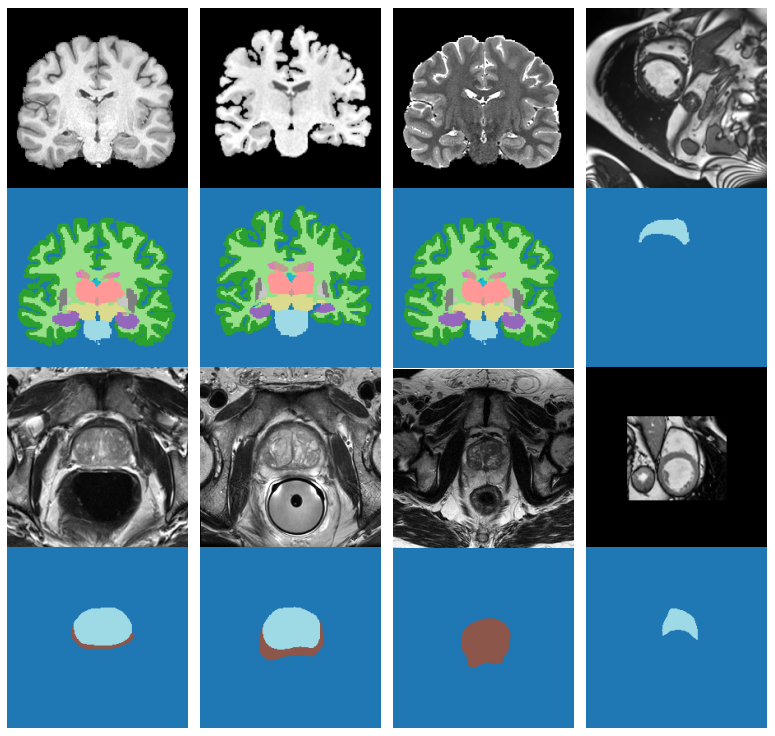}
\caption{Figure shows examples of images from the datasets used in our experiments, 3 brain, 3 prostate and 2 cardiac datasets. Beneath each image we show the corresponding ground truth segmentation map. For the prostate datasets, the first two datasets show labels for two prostate sub-glands separately, while the third dataset contains labels for the entire prostate gland. Note that, for each anatomy, while the underlying organ structure remains relatively consistent across domains (as can be seen by the similarity of the ground truth segmentations), the images acquired by different scanners / protocols have varying contrasts.}\label{fig_illustrate_domain_shift}
\end{figure}

\vspace{0.1cm} \noindent Fig.~\ref{fig_illustrate_domain_shift} shows some images from the different datasets and their corresponding segmentations. Note that, for each anatomy, the underlying segmentations are very similar across domains, while the image contrasts vary.  Table~\ref{tab:dataset_details} shows our training, test and validation split (in terms of number of 3D images) for each dataset.

\begin{table}
    \small
    \caption{Dataset Details}~\label{tab:dataset_details}
    
    \centering
    \begin{tabular}{|l|l|l|l|l|l|}
    
        \hline
        \rowcolor{titlegray} 
        Anatomy & Dataset & SD / TD & $N_{train}$ & $N_{val}$ & $N_{test}$\\
        \hline
        Brain & HCP-T$_{1}$ & SD$_{}$ & 20 & 5 & 20\\
        Brain & ABIDE-T$_{1}$ & TD$_{1}$ & 10 & 5 & 20\\
        Brain & HCP-T$_{2}$ & TD$_{2}$ & 20 & 5 & 20\\
        \hline
        Prostate & NCI & SD$_{}$ & 15 & 5 & 10\\
        Prostate & USZ & TD$_{1}$ & 28 & 20 & 20\\
        Prostate & PROMISE & TD$_{2}$ & 20 & 10 & 20\\
        \hline
        Heart & ACDC & SD$_{}$& 120 & 40 & 40\\
        Heart & RVSC & TD$_1$ & 48 & 24 & 24\\
        \hline
        
    \end{tabular}
    
\end{table}

\subsection{Pre-processing}\label{sec:preprocessing}
\noindent We pre-process all images with the following steps.
Firstly, we remove any bias fields with the N4 algorithm~\cite{tustison2010n4itk}.
Secondly, we carry out $0-1$ intensity normalization per image as: $x_{normalized} = (x - x_{p}^{1}) / (x_{p}^{99} - x_{p}^{1})$, where $x_{p}^{i}$ denotes the $i^{th}$ percentile of the intensity values in the image volume, followed by clipping the intensities at 0 and 1.
For the brain datasets, this is followed by skull stripping, setting intensities of all non-brain voxels to 0. 

\vspace{0.1cm} \noindent We train segCNN in 2D (due to GPU memory limitations), and the DAE in 3D (in order to exploit 3D organ structure). For the 2D segCNN, we rescale all images to fixed pixel-size in the in-plane dimensions followed by cropping and / or padding with zeros to match the image sizes to a fixed size for each anatomy. The fixed pixel-sizes for the brain, prostate and cardiac datasets are 0.7mm$^2$, 0.625mm$^2$ and 1.33mm$^2$ respectively, while the fixed image size is 256x256 for all anatomies. The ground truth labels of the training and validation images are rescaled and cropped / padded in the same way as the corresponding images. Test images are also rescaled and cropped / padded before predicting their segmentations. The predicted segmentations, however, are rescaled back and evaluated in their original pixel-size to avoid any experimental biases.

\vspace{0.1cm} \noindent For the 3D DAE, we pre-process the segmentation labels with rescaling and cropping / padding applied in all 3 dimensions. The fixed voxel-sizes are set to 2.8x0.7x0.7mm$^3$, 2.5x0.625x0.625mm$^3$ and 5.0x1.33x1.33mm$^3$ for the brain, prostate and cardiac datasets, respectively, while the fixed 3D image size is set to 64x256x256 for the brain images and 32x256x256 for the other two anatomies.

\subsection{Implementation Details}\label{sec:implementation_details}
\subsubsection{Network Architectures}
\noindent We implement the image normalization CNN, $N_{\phi}$, with $n_N = 3$ convolutional layers, with the respective number of output channels set to 16, 16 and 1, each using kernels of size $n_k = 3$. Keeping in mind the relatively small depth of $N_{\phi}$, we equip it with an expressive activation function, $act(x) = exp(-x^2 / \sigma^2)$, where the scale parameter $\sigma$ is trainable and different for each output channel.

\vspace{0.1cm} \noindent For modeling the normalized-image-to-segmentation CNN, $S_{\theta}$, as well as the DAE, $D_{\psi}$, we use an encoder-decoder architecture with skip connections across corresponding depths, in spirit of the commonly used U-Net~\cite{ronneberger2015u} architecture.
Batch normalization~\cite{ioffe2015batch} and the ReLU activation function~\cite{nair2010rectified} are used in both networks.
Bilinear upsampling is preferred to deconvolutions in light of the potential checkerboard artifacts while using the latter~\cite{odena2016deconvolution}. That said, we would like to emphasize that the proposed test-time adaptation strategy, the normalization module and the DAE are agnostic to the architecture of the normalized-image-to-segmentation CNN. Any architecture can be used instead of the U-Net architecture we used in our experiments.

\subsubsection{Optimization Details}
\noindent We use the Dice loss~\cite{milletari2016v} as the training loss function for all networks. The batch size is set to $16$ for the 2D segmentation CNN training and the test-time adaptation, and to $1$ for the 3D DAE. We use the Adam optimizer~\cite{kingma2014adam} with default parameters and a learning rate of $10^{-3}$. We train the segmentation CNN and DAE for $50000$ iterations and chose the best models based on validation set performance.

\vspace{0.1cm} \noindent For the test-time adaptation for each test image, we run the inference-time optimization for $T=500$ gradient updates for the brain datasets and for $T=7500$ gradient updates for the other two anatomies (see Sec.~\ref{sec:gradient_accumulation}, step 2).\footnote{This discrepancy is due to the differences in the number of slices of the datasets. In our current implementation, each update of the parameters $\phi$ is performed with an average gradient over 16 batches for the brain datasets and over 2 batches for the prostate and cardiac datasets. To account for lower number of batches for the latter, we use larger number of gradient update steps. Thus, effectively, even with the different number of gradient updates, images from all datasets observe roughly the same number of batches during the optimization.} The denoised labels that are used to drive the optimization are updated every $f=25$ steps (see Sec.~\ref{sec:gradient_accumulation}, step 1). During the update iterations, parameters that lead to the highest Dice score between the DAE input and output are chosen as optimal for a given test image. Additionally, we run a separate 'fast' version of our method, where we carry out the test-time adaptation with the aforementioned hyper-parameters for the first test image of each TD. For subsequent images of that TD, we initialize the parameters of the normalization module with the optimal parameters corresponding to the first TD image. This provides a better starting point for the optimization, so we run it for $T=100$ gradient updates for the brain datasets and for $T=1500$ gradient updates for the other anatomies. On a NVIDIA GeForce GTX TITAN X GPU, the test time adaptation requires about 1 hour for the first image of a particular TD and about 12 minutes for each image thereafter with our experimental implementation, which could be further optimized for time efficiency.

\subsubsection{Data Augmentation}~\label{sec:data_aug}
\noindent In~\cite{zhang2019generalizing}, data augmentation has been shown to be highly effective for improving cross-scanner robustness in medical image segmentation. Accordingly, we train segCNN with a suite of data augmentations, consisting of geometric as well as intensity transformations. As geometric transformations, we use translation ($\sim U(-10, 10)$ pixels), rotation ($\sim U(-10, 10)$ degrees), scaling ($\sim U(0.9, 1.1)$) and random elastic deformations (obtained by generating random noise images between $-1$ and $1$, smoothing them with a Gaussian filter with standard deviation $20$ and scaling them with a factor of $1000$)~\cite{simard2003best}. As intensity transformations, we use gamma transformation ($x_{aug} = x^c; c \sim U(0.5, 2.0)$), brightness changes ($x_{aug} = x + b; b \sim ~U(0.0, 0.1)$) and additive Gaussian noise ($x_{aug}^{ij} = x^{ij} + n^{ij}; n^{ij} \sim ~N(0.0, 0.1)$, where the superscript $ij$ is used to indicate that the noise is added independently for each pixel in the image). For the cardiac datasets, we observe that the images are acquired in different orientations, so for this anatomy, we add to the set of geometric transformations: rotations by multiple of $90$ degrees and left-right and up-down flips. The geometric transformations are applied to both images as well as segmentations, while the intensity transformations are applied only to the images. We also train the DAE with data augmentation consisting of geometric transformations applied on the segmentations.
Each transformation is applied with a probability of $0.25$ to each image in a training mini-batch.

\subsubsection{Noise Hyper-parameters for DAE Training}
\noindent We visually inspected the generated corrupted segmentations by using different noise hyper-parameters. Based on this, we chose the maximum number of patches to be copied, $n_1^{max} = 200$ and the maximum size of a patch, $n_2^{max} = 20$. During its training, we determined the best DAE model based on its denoising performance on a corrupted validation dataset, which we generate by corrupting each validation image 50 times with the noising process described in Sec.~\ref{sec:dae_training}. 
The validation dataset used to select the best DAE model is a part of the SD, not TD. 

\subsubsection{Hyper-parameters for atlas-based initial optimization in case of domain shifts involving a protocol change}
\noindent For the brain datasets, the SD consists of T1w images, while the TD$_{2}$ consists of T2w images, both from the HCP dataset, but from different subjects. In this case, we used the atlas based initial optimization described in Section~\ref{sec:atlas_initial_training}.
As the images in the HCP dataset are already rigidly registered, we create an atlas by converting the SD labels to one-hot representations and averaging them voxel-wise. To decide when the optimization switches from being driven by the atlas to the DAE, we use the thresholding-based method described in Sec.~\ref{sec:atlas_initial_training}, setting the hyper-parameters $\alpha = 1.0$ and $\beta = 0.25$.

\subsubsection{Evaluation Criteria}
\noindent We evaluate the predicted segmentations by comparing them with corresponding ground truth segmentations using the Dice coefficient~\cite{dice1945measures} and the 95$^{th}$ percentile of Hausdorff distance~\cite{huttenlocher1993comparing}. We report mean values of these scores across foreground labels, all test images and across 3 runs of each experiment.

\subsection{Results}\label{sec:results}
\begin{table*}[h!]
    \begin{adjustwidth}{-.5in}{-.5in}  
    \small
    \caption{Quantitative Results: The top and the bottom table show Dice scores and 95$^{th}$ percentile Hausdorff distance respectively. Mean results over all foreground labels, all test subjects and over 3 experiment runs are shown.
    DA stands for Data Augmentation and TTA for Test Time Adaptation.
    For brain, SD: HCP-T1w, TD$_1$: ABIDE-Caltech-T1w, TD$_2$: HCP-T2w, for prostate, SD: NCI, TD$_1$: USZ, TD$_2$: PROMISE12, for heart, SD: ACDC, TD$_1$: RVSC.
    For the prostate datasets, results are shown for whole gland segmentation (whole) as well as averaged over the central gland and peripheral zone (sep.). The rows show different training / adaptation strategies and are described in the text. The best results for the TDs among the various domain generalization methods are in boldface. For the proposed method, the * next to Dice scores denotes statistical significance over 'SD + DA + Post-Proc.' (Permutation test with a threshold value of 0.01). }~\label{tab:quant_results1}
    
    \centering
    \begin{tabular}{| l| *{3}{c|} *{3}{c|} *{2}{c|} *{2}{c|} }
    
        \hline
        \rowcolor{titlegray} 
        Anatomy & \multicolumn{3}{c|}{Brain} & \multicolumn{3}{c|}{Prostate (whole)} & \multicolumn{2}{c|}{Prostate (sep.)} & \multicolumn{2}{c|}{Heart} \\\hline
        
        \rowcolor{titlegray} 
        \backslashbox{Train}{Test} & SD & TD$_1$ & TD$_2$ & SD & TD$_1$ & TD$_2$ & SD & TD$_1$ & SD & TD$_1$
        \\\hline
        
        \rowcolor{lightgray} \multicolumn{11}{|l|}{{{Baseline and Benchmark}}} \\ \hline
        
        \rowcolor{lightergray} 
        SD (Baseline) & 
        0.853 & 
        0.588 & 
        0.107 & 
        0.840 & 
        0.586 & 
        0.609 & 
        0.722 & 
        0.544 & 
        0.823 & 
        0.670 
        \\\hline
        
        \rowcolor{lightergray} 
        TD$_n$ (Benchmark) & 
        - & 
        0.896 & 
        0.867 & 
        - & 
        0.817 & 
        0.834 & 
        - & 
        0.732 & 
        - & 
        0.806 
        \\\hline
        
        \rowcolor{lightgray} \multicolumn{11}{|l|}{{{Relevant Methods}}} \\ \hline

        \rowcolor{lightergray}
        MLDG [\cite{li2018learning}] &
        0.874 & 
        0.686 & 
        0.074 & 
        0.913 & 
        0.772 & 
        0.756 & 
        0.818 & 
        0.658 & 
        0.844 & 
        0.696 
        \\\hline
        
        \rowcolor{lightergray}
        MASF [\cite{dou2019domain}] &
        0.870 & 
        0.693 & 
        0.073 & 
        0.913 & 
        0.751 & 
        0.781 & 
        0.817 & 
        0.640 & 
        0.838 & 
        0.698 
        \\\hline
        
        \rowcolor{lightergray}
        MLDGTS [\cite{zhang2020generalizable}] &
        0.876 & 
        0.733 & 
        0.072 & 
        0.912 & 
        0.711 & 
        0.761 & 
        0.815 & 
        0.608 & 
        0.831 & 
        0.361 
        \\\hline
        
        \rowcolor{lightergray}
        SD $+$ DA [\cite{zhang2019generalizing}] (Strong baseline) & 
        0.876 & 
        0.753 & 
        0.083 & 
        0.911 & 
        0.769 & 
        0.786 & 
        0.815 & 
        0.656 & 
        0.834 & 
        0.744 
        \\\hline
        
        \rowcolor{lightergray}
        \begin{tabular}{@{}l@{}} SD $+$ DA $+$ Post-Proc. [\cite{larrazabal2019anatomical}] \\ \end{tabular} & 
        - & 
        0.706 & 
        0.112 & 
        - & 
        0.789 & 
        0.823 & 
        - & 
        0.678 & 
        - & 
        0.746 
        \\\hline
        
        \rowcolor{lightgray} \multicolumn{11}{|l|}{{{\textbf{Proposed Method}}}} \\ \hline
        
        \rowcolor{lightergray}
        SD $+$ DA $+$ TTA (Adapt $\phi$, using DAE) & 
        - & 
        \textbf{0.800}$^*$ & 
        \textbf{0.733}$^*$ & 
        - & 
        0.790 & 
        \textbf{0.858}$^*$ & 
        - & 
        0.676 & 
        - & 
        0.742 
        \\\hline
        
        \rowcolor{lightergray}
        SD $+$ DA $+$ TTA (Adapt $\phi$, using DAE) - \textbf{Fast} & 
        - & 
        \textbf{0.800} & 
        0.728 & 
        - & 
        0.790 & 
        0.842 & 
        - & 
        \textbf{0.683} & 
        - & 
        \textbf{0.747} 
        \\\hline
        
        \rowcolor{lightgray} \multicolumn{11}{|l|}{{{Ablation Studies}}} \\ \hline

        \rowcolor{lightergray}
        \begin{tabular}{@{}l@{}} SD $+$ DA $+$ TTA (Adapt $\phi, \theta$, using DAE) \end{tabular} & 
        - & 
        0.671 & 
        0.650 & 
        - & 
        0.718 & 
        0.606 & 
        - & 
        0.583 & 
        - & 
        0.713 
        \\\hline
        
        \rowcolor{lightergray}
        \begin{tabular}{@{}l@{}} SD $+$ DA $+$ TTA (Adapt $\phi$, using GT labels) \end{tabular} & 
        - & 
        0.831 & 
        0.837 & 
        - & 
        0.836 & 
        - & 
        - & 
        0.771 & 
        - & 
        - 
        \\\hline
        
        \rowcolor{lightergray}
        SD $+$ DA $+$ Post.Proc. 10 passes through DAE & 
        - & 
        0.633 & 
        0.106 & 
        - & 
        \textbf{0.791} & 
        0.826 & 
        - & 
        \textbf{0.683} & 
        - & 
        0.731 
        \\\hline
        
        \rowcolor{lightergray}
        SD $+$ DA $+$ Post.Proc. 100 passes through DAE & 
        - & 
        0.529 & 
        0.101 & 
        - & 
        0.789 & 
        0.823 & 
        - & 
        0.672 & 
        - & 
        0.688 
        \\\hline

    \end{tabular}
    
    \vspace{0.1cm}
    
    \centering
    \begin{tabular}{| l| *{3}{c|} *{3}{c|} *{2}{c|} *{2}{c|} }
    
        \hline
        \rowcolor{titlegray}
        Anatomy & \multicolumn{3}{c|}{Brain} & \multicolumn{3}{c|}{Prostate (whole)} & \multicolumn{2}{c|}{Prostate (sep.)} & \multicolumn{2}{c|}{Heart} \\\hline
        
        \rowcolor{titlegray}
        \backslashbox{Train}{Test} & SD & TD$_1$ & TD$_2$ & SD & TD$_1$ & TD$_2$ & SD & TD$_1$ & SD & TD$_1$
        \\\hline
        
        \rowcolor{lightgray} \multicolumn{11}{|l|}{{{Baseline and Benckmark}}} \\ \hline
        
        \rowcolor{lightergray}
        SD (Baseline) & 
        9.11 & 
        34.34 & 
        52.13 & 
        16.65 & 
        147.88 & 
        68.97 & 
        19.09 & 
        146.25 & 
        15.89 & 
        16.69 
        \\\hline
        
        \rowcolor{lightergray}
        TD$_n$ (Benchmark) & 
        - & 
        1.42 & 
        2.32 & 
        - & 
        21.91 & 
        14.25 & 
        - & 
        22.35 & 
        - &  
        3.52 
        \\\hline
        
        \rowcolor{lightgray} \multicolumn{11}{|l|}{{{Relevant Methods}}} \\ \hline
        
        \rowcolor{lightergray}
        MLDG [\cite{li2018learning}] &
        2.49 & 
        21.22 & 
        52.28 & 
        2.51 & 
        43.53 & 
        32.34 & 
        4.65 & 
        42.67 & 
        7.74 & 
        22.49 
        \\\hline
        
        \rowcolor{lightergray}
        MASF [\cite{dou2019domain}] &
        2.13 & 
        18.45 & 
        57.90 & 
        5.60 & 
        90.15 & 
        45.79 & 
        6.56 & 
        87.32 & 
        6.20 & 
        15.92 
        \\\hline
        
        \rowcolor{lightergray}
        MLDGTS [\cite{zhang2020generalizable}] &
        2.19 & 
        13.52 & 
        55.00 & 
        3.58 & 
        93.93 & 
        36.56 & 
        6.50 & 
        92.07 & 
        8.41 & 
        42.79 
        \\\hline
        
        \rowcolor{lightergray}
        SD $+$ DA [\cite{zhang2019generalizing}] (Strong Baseline) & 
        2.08 & 
        17.99 & 
        55.81 & 
        3.59 & 
        55.83 & 
        26.92 & 
        5.25 & 
        53.31 & 
        6.23 & 
        13.90 
        \\\hline
        
        \rowcolor{lightergray}
        SD $+$ DA $+$ Post-Proc. [\cite{larrazabal2019anatomical}] & 
        - & 
        13.16 & 
        53.58 & 
        - & 
        38.86 & 
        14.31 & 
        - & 
        38.52 & 
        - & 
        9.46 
        \\\hline
        
        \rowcolor{lightgray} \multicolumn{11}{|l|}{{{\textbf{Proposed Method}}}} \\ \hline
        
        \rowcolor{lightergray}
        SD $+$ DA $+$ TTA (Adapt $\phi$, using DAE) & 
        - & 
        10.14 & 
        21.51 & 
        - & 
        28.08 & 
        \textbf{9.50} & 
        - & 
        31.42 & 
        - & 
        15.29 
        \\\hline
        
        \rowcolor{lightergray}
        SD $+$ DA $+$ TTA (Adapt $\phi$, using DAE) - \textbf{Fast} & 
        - & 
        9.21 & 
        19.18 & 
        - & 
        30.76 & 
        9.82 & 
        - & 
        33.85 & 
        - & 
        12.66 
        \\\hline
        
        \rowcolor{lightgray} \multicolumn{11}{|l|}{{{Ablation Studies}}} \\ \hline
        
        \rowcolor{lightergray}
        \begin{tabular}{@{}l@{}} SD $+$ DA $+$ TTA (Adapt $\phi, \theta$, using DAE) \end{tabular} & 
        - & 
        \textbf{4.99} & 
        \textbf{13.69} & 
        - & 
        \textbf{15.49} & 
        16.09 & 
        - & 
        \textbf{22.24} & 
        - & 
        \textbf{4.83} 
        \\\hline
        
        \rowcolor{lightergray}
        \begin{tabular}{@{}l@{}} SD $+$ DA $+$ TTA (Adapt $\phi$, using GT labels) \end{tabular} & 
        - & 
        6.12 & 
        3.88 & 
        - & 
        35.20 & 
        - & 
        - & 
        35.35 & 
        - & 
        - 
        \\\hline
        
        \rowcolor{lightergray}
        SD $+$ DA $+$ Post.Proc. 10 passes through DAE & 
        - & 
        15.14 & 
        56.48 & 
        - & 
        28.54 & 
        11.56 & 
        - & 
        30.81 & 
        - & 
        7.23 
        \\\hline
        
        \rowcolor{lightergray}
        SD $+$ DA $+$ Post.Proc. 100 passes through DAE & 
        - & 
        21.57 & 
        58.26 & 
        - & 
        25.72 & 
        10.38 & 
        - & 
        29.27 & 
        - & 
        9.24 
        \\\hline

    \end{tabular}
    
    \end{adjustwidth}
\end{table*}

\begin{figure*}[h!]
\centering
    \includegraphics[trim = 0mm 0mm 0mm 0mm, angle=0, clip, width=1.0\textwidth]{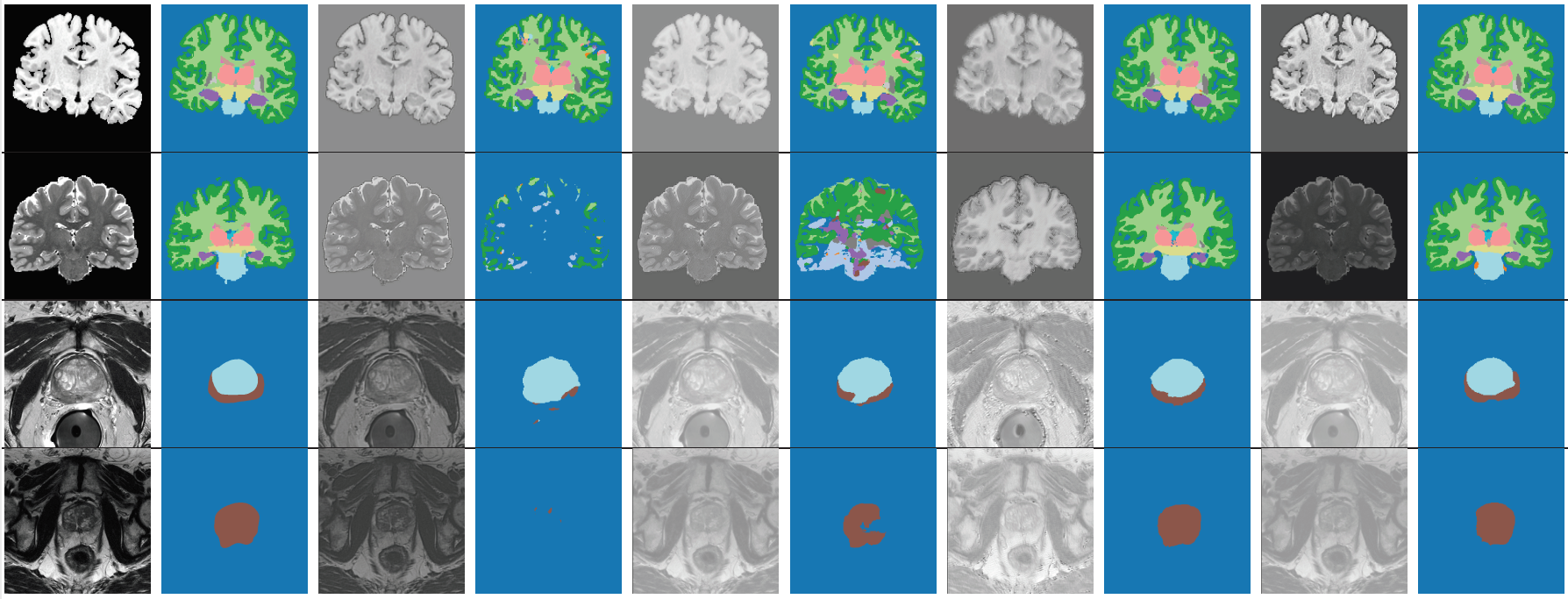}
\caption{Qualitative results. Rows show results from TDs for different anatomies. The first and second columns show test images and their ground truth segmentation respectively. After this, from left to right, normalized image and predicted segmentation pairs are shown for training with SD, SD + DA, SD + DA + TTA and TD.  }\label{fig_qualitative_results}
\end{figure*}

\noindent Table~\ref{tab:quant_results1} shows the results of our experiments.
In the following sub-sections, we describe the different compared methods.

\subsubsection{Baseline and Benchmark}
\noindent Firstly, for each anatomy, we trained a segmentation CNN on the SD and tested it on the TDs. This is shown in the first row of Table~\ref{tab:quant_results1} and provides a baseline performance for the problem.
Next, we trained specialized segmentation CNNs for each TD, using a separate training and validation set from that TD. The performance of such specialized CNNs forms the benchmark for the problem (for the purposes of this work) and is reported in the second row of Table~\ref{tab:quant_results1}.
The difference between these two rows shows the gap in generalization performance that domain generalization methods seek to fill.

\subsubsection{Relevant Methods}
\noindent Next, we evaluated state-of-the-art domain generalization methods proposed in the literature.

\vspace{0.1cm} \noindent \newtextRtwo{
\textit{4.4.2.1. Meta Learning based Methods}:
Several meta learning based approaches~\cite{li2018learning, dou2019domain, zhang2020generalizable} have been proposed for tackling the domain generalization problem. The main idea of such methods is to simulate the domain shift problem during the training of the segmentation CNN. This is done by having meta-train and meta-test domains during training and requiring that the gradient updates for the meta-train domains be such that the task loss is also minimized on the meta-test domains. As we only had access to a single SD for training, we simulated meta-train and meta-test domains by using different gamma transformations in each batch of training. Additionally, we used all other data augmentation transformations described in Sec.~\ref{sec:data_aug}.
We observed that all meta learning based approached substantially improved the domain generalization performance over the baseline, for cases where the source and target domain images are acquired with the same imaging protocol.}

\vspace{0.1cm} \noindent \textit{4.4.2.2. Data Augmentation}:
\cite{zhang2019generalizing} show that extensive data augmentation (DA) by stacking different transformations greatly improves the segmentation performance on unseen scanners and protocols.
We also observed a remarkable performance boost due to data augmentation for the cases where the source and target domain images are acquired with the same imaging protocol.
Nonetheless, there still remains a gap with respect to training separately on the TDs.
We refer to the training with data augmentation as a \textit{strong baseline} that we seek to improve upon with our method.

\vspace{0.1cm} \noindent \textit{4.4.2.3. Post-Processing with DAEs}:
\cite{larrazabal2019anatomical} propose post-processing the predicted segmentations with DAEs in order to increase their plausibility.
We use our trained DAEs in order to carry out such post-processing on the segmentations predicted by the 'SD + DA' trained CNNs.
From Table~\ref{tab:quant_results1}, it can be observed that the post-processing brought about substantial improvements over 'SD + DA', especially for the prostate dataset.
However, the DAE post-processing lead to performance degradation on the brain datasets.
We believe that this can be attributed to the fact that DAEs map their inputs to a plausible segmentation, however, that segmentation may not necessarily be tied to the test image.
Furthermore, the post-processing method did not work when the source and target domains are acquired with different protocols or imaging modalities. In such cases, the pre-trained segCNN predicted highly corrupted segmentations, which cannot be seen as samples from the DAE's training input distribution. Therefore, post-processing with the trained DAE could not improve the segmentation accuracy.
This can be seen for the brain datasets, when the target domain is TD$_2$, which consists of T2w images, while the SD consists of T1w images.

\subsubsection{Test-Time Adaptation}
\noindent For the proposed method, we first trained segCNN on the SD along with DA. Then, we adapted the I2NI CNN, $N_\phi$, for each test image, according to the proposed framework. It can be seen that the test-time adaptation provided substantial performance gains over competing methods across all datasets for brain and prostate anatomies. Qualitative results shown in Fig~\ref{fig_qualitative_results} reveal similarly substantial improvements over 'SD + DA', especially for the brain and prostate datasets. It can be seen that the test-time adaptation improved the predicted segmentation by, for instance, correcting predictions that are contextually misplaced, completing organ shapes and removing outliers. For the cardiac dataset, we observed that the SD training along with data augmentation already provided a fairly good segmentation. The proposed method preserved this performance, but could not further improve it. The improvement in Dice scores with the proposed method as compared to post-processing using the trained DAEs (SD + DA + Post-Proc.) was statistically significant for 3 out of the 5 datasets (marked with * in the Table), as measured using a paired Permutation test with 100000 permutations. For the other datasets, we obtained similar results upon direct post-processing as with the test-time adaptation using DAEs.

\vspace{0.1cm} \noindent We also carried out a 'fast' version of our method, where the optimal I2IN parameters for the first test image of each TD were used for initialization in the test-time adaptation for subsequent images of that TD.
\newtextRtwo{After such improved initialization, the adaptation for subsequent images could be completed in fewer iterations, thus reducing the time required for the adaptation.
However, initializing the I2NI parameters from values obtained from the SD training lead to similar results as the new initialization strategy.
This demonstrates that the proposed adaptation works in a stable manner for multiple test images, and is not dependent on a 'good' first test image from a new TD. In practice, the fast version is more attractive due to the time saving.}

\newtextRtwosecondreview{
\subsubsection{Comparison with Unsupervised Domain Adaptation}
\noindent Unsupervised Domain Adaptation (UDA) is widely proposed in the literature for tackling the domain shift problem.
In this section, we compare the performance of the proposed test-time adaptation (TTA) method with UDA works.
Note that UDA methods work in a more relaxed setting, where the labelled SD dataset is assumed to be available while adapting for each new TD. Although it may be challenging to meet this requirement in practice, we carry out this experiment to quantify the potential advantages of such approaches.
Table~\ref{tab:uda_comparison} shows the results of these experiments.
Typically, UDA methods utilize a set of unlabelled images from the TD, $X_{TD}^{tr}$, for the adaptation, but use a separate set of test images from the TD, $X_{TD}^{ts}$,  for evaluation. However, as ground truth labels of $X_{TD}^{tr}$ are not utilized for the adaptation, we present evaluations for this set as well, which was also proposed as a more suitable UDA setting in a recent work~\cite{varsavsky2020testtime}. The size of the image sets $X_{TD}^{tr}$ and $X_{TD}^{ts}$ was the same as specified in Table~\ref{tab:dataset_details}.

\vspace{0.1cm} \noindent We conducted experiments with two representative UDA methods: 1)~\cite{kamnitsas2017unsupervised}, where an adversarial loss is employed to incentivize invariance in the SD and TD features, and 2)~\cite{huo2018synseg}, where a transformation network between the TD and the SD is trained, and then the transformed images are passed through the segmentation network.
Despite our persistent efforts, application of the method by~\cite{kamnitsas2017unsupervised} to large domain changes (modality change) and of the method by~\cite{huo2018synseg} to small domain changes (scanner changes within the same modality) led to poorer accuracies that the SD + DA baseline.
To the best of our knowledge, this observation is consistent with the domain shifts in the datasets presented in these works as well as other UDA works that follow similar ideas.

\vspace{0.1cm} \noindent Remarkably, it can be seen that not using SD labelled dataset does not hinder the TTA method; it achieves comparable results to the best performing UDA methods, especially for the scanner change related domain shifts.
The image-to-image translation method using cycleGAN~\cite{huo2018synseg} provided better performance than TTA for the case with large domain shift (modality change), but lead to poorer results than the baseline for smaller domain shifts (scanner changes).
It is also noteworthy that within the UDA methods, the performance gains for $X_{TD}^{tr}$ are similar to that for $X_{TD}^{ts}$. Thus, using the same images for adaptation as well as testing did not lead to additional improvements in our experiments.
These results highlight that test-time adaptation can serve as a potent and more flexible alternative to UDA methods, for both small and large domain changes.}

\begin{table*}[t!]
    \small
    \caption{\newtextRtwothirdreview{Dice scores} for comparison of the proposed method with unsupervised domain adaptation (UDA) methods. These results show that the proposed test-time adaptation method performs comparably with the best performing UDA methods, with the additional critical benefit of not requiring the labelled SD dataset while adapting for each new TD.
    }~\label{tab:uda_comparison}
    
    \centering
    \begin{tabular}{| l| *{4}{c|} *{4}{c|} *{2}{c|} }
    
        \hline
        \rowcolor{titlegray} 
        Anatomy & \multicolumn{4}{c|}{Brain} & \multicolumn{4}{c|}{Prostate (whole)} &  \multicolumn{2}{c|}{Heart} \\\hline
        
        \rowcolor{titlegray} 
        \backslashbox{Train}{Test} &
        TD$_1^{tr}$ & TD$_1^{ts}$ &
        TD$_2^{tr}$ & TD$_2^{ts}$ &
        TD$_1^{tr}$ & TD$_1^{ts}$ &
        TD$_2^{tr}$ & TD$_2^{ts}$ &
        TD$_1^{tr}$ & TD$_1^{ts}$
        \\\hline
        
        \rowcolor{lightergray}
        SD $+$ DA [\cite{zhang2019generalizing}] (Strong baseline) & 
        0.750 & 
        0.753 & 
        0.081 & 
        0.083 & 
        0.767 & 
        0.769 & 
        0.747 & 
        0.786 & 
        0.715 & 
        0.744 
        \\\hline
        
        \rowcolor{lightergray}
        UDA - Invariant features [\cite{kamnitsas2017unsupervised}] &
        0.792 & 
        0.798 & 
        \newtextRtwothirdreview{0.081} & 
        \newtextRtwothirdreview{0.083} & 
        0.789 & 
        0.793 & 
        0.766 & 
        0.802 & 
        0.724 & 
        0.750 
        \\\hline
        
        \rowcolor{lightergray}
        UDA - Image-to-Image translation [\cite{huo2018synseg}] &
        \newtextRtwothirdreview{0.646} & 
        \newtextRtwothirdreview{0.639} & 
        0.816 & 
        0.813 & 
        \newtextRtwothirdreview{0.607} & 
        \newtextRtwothirdreview{0.694} & 
        \newtextRtwothirdreview{0.765} & 
        \newtextRtwothirdreview{0.747} & 
        \newtextRtwothirdreview{0.252} & 
        \newtextRtwothirdreview{0.167} 
        \\\hline
        
        \rowcolor{lightergray}
        TTA (Proposed) & 
        - & 
        0.800 & 
        - & 
        0.733 & 
        - & 
        0.790 & 
        - & 
        0.858 & 
        - & 
        0.742 
        \\\hline

        \rowcolor{lightergray} 
        TD$_n$ (Benchmark) & 
        - & 
        0.896 & 
        - & 
        0.867 & 
        - & 
        0.817 & 
        - & 
        0.834 & 
        - & 
        0.806 
        \\\hline
        
    \end{tabular}
    
\end{table*}

\subsubsection{Analysis experiments}
\noindent Finally, we conducted several experiments to analyze the importance of the design choices in the proposed method. The results of these experiments are in the lower rows of Table~\ref{tab:quant_results1}.

\vspace{0.1cm} \noindent i. Firstly, we studied the importance of restricting the adaptation to just the the I2NI part of segCNN. That is, we trained segCNN on the SD along with DA and adapted all its parameters, \{$\phi$, $\theta$\}, for each test image, according to the proposed framework.
It can be seen that this lead to a drop in segmentation accuracy in terms of Dice score, but improved the Hausdorff distance (in all TDs, except TD$_2$ for prostate). Qualitatively, we observed that the Dice scores deteriorated because the segmentations while becoming more plausible, became inaccurate around the edges. The Hausdorff distance, on the other hand, improved because the added flexibility allowed outliers to be removed more effectively. Overall, we believe that accurate segmentations around organ edges are more valuable than removing extreme outliers (which can be removed by other post-processing steps, if required). Thus, we believe that this experiment showcases the importance of freeze a majority of the parameters at the values obtained from the initial supervised learning.

\vspace{0.1cm} \noindent ii. Secondly, we examined if the flexibility afforded by the adaptable I2NI CNN is sufficient for obtaining accurate segmentations via test-time adaptation. To this end, we trained segCNN using SD along with DA, and then adapted the I2NI CNN for each test image, driving the test-time adaptation using the ground truth labels of the test image. This is an ablation study that removed the DAE from the picture and asked the following question: if an oracle were available to drive the test-time adaptation, can the I2NI CNN be appropriately adapted to follow the oracle?
This experiment was not done for the PROMISE and the RVSC datasets as the annotations for these datasets are for a different set of organs / tissues as compared to the corresponding SD datasets.
As expected, we observed improvements in accuracy in all cases, as compared to the 'SD + DA' baseline. However, it is interesting to note that for the brain TD datasets, the resulting accuracies were inferior to those obtained by training segCNN separately for each TD. This shows that despite the test-time adaptation, some bias towards the SD may remain in the NI2S CNN.

\begin{figure}[h!]
\centering
    \includegraphics[trim = 0mm 0mm 0mm 0mm, angle=0, clip, width=0.49\textwidth]{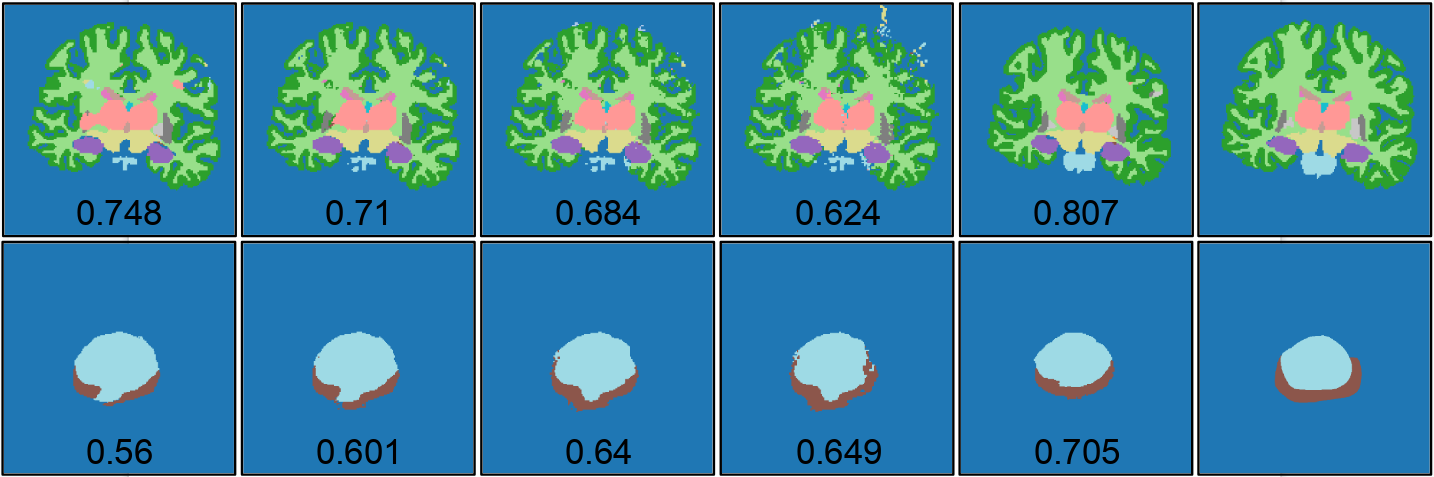}
\newtextRone{\caption{Multiple passes through the DAE do not suffice to improve segmentation accuracy. From left to right: Initial prediction by segCNN trained on SD+DA, followed by 1, 10 and 100 passes through the DAE, followed by the results of the test-time adaptation (SD + DA + TTA) and finally, the ground truth. Top and bottom rows show results for the ABIDE-Caltech and USZ datasets, respectively, and the numbers below the segmentations are the corresponding volumetric averaged foreground Dice scores.}\label{fig_dae_multiple_passes}}
\end{figure}

\vspace{0.1cm} \noindent \newtextRone{iii. Lastly, we asked the question: Is test-time adaptation required at all, or can the accuracy on TDs be improved by simply passing the predicted segmentation multiple times through the DAE? The results of this experiment are shown in last two rows of Table~\ref{tab:quant_results1} and in Fig.~\ref{fig_dae_multiple_passes}, for different number of passes through the DAE. We observed that such a post-processing approach could not improve segmentation accuracy as much as the test-time adaptation. On the contrary, for the brain datasets, where the segmentations are more complicated than other anatomies due to the presence of multiple structures, the post-processing with multiple DAE passes worsened the segmentation accuracy. We believe that this might be because the DAE output, although generally plausible according the labels in the SD dataset, is not necessarily tied to the input image in question. The proposed method constrains the predicted segmentations to be tied to the input image by: (1) freezing the NI2S CNN and (2) keeping the adaptable I2NI CNN relatively shallow. Also, the limited flexibility for the adaptation guards against potential errors of the DAE such as the one seen fourth column for the brain dataset in Fig.~\ref{fig_dae_multiple_passes}.}

\begin{figure}[h!]
\centering
    \includegraphics[trim = 0mm 0mm 0mm 0mm, angle=0, clip, width=0.49\textwidth]{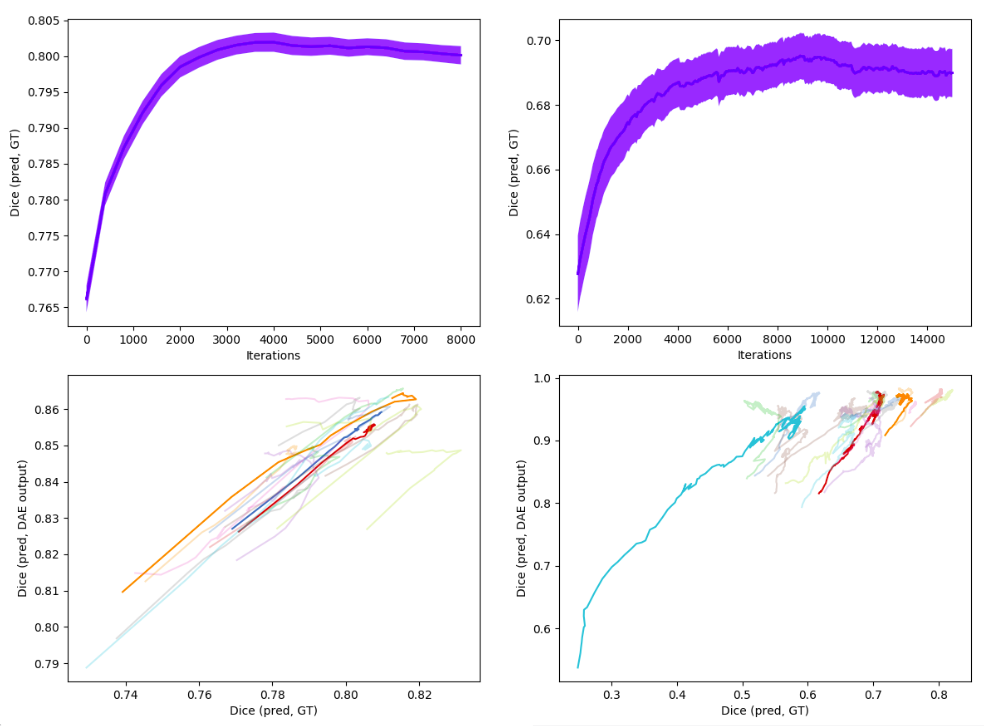}
\newtextRone{\caption{Convergence of the test-time adaptation. Columns are for different TDs - ABIDE-Caltech (left), USZ (right). The top row shows the mean and 0.1 * std. deviation over different test subjects, of the Dice between the predicted and ground truth segmentation, as a function of the test-time adaptation iterations. The bottom row shows the correlation between (a) the Dice between the predicted and ground truth segmentation and (b) the Dice between the DAE input and DAE outputs. Each color represents the test-time adaptation of a single test subject. For clarity, a few subjects are shown in opaque colours, while others are faded.}\label{fig_convergence_results}}
\end{figure}

\newtextRone{
\subsubsection{Convergence of Test-Time Adaptation}
\noindent We note that the convergence of the test-time adaptation is not theoretically guaranteed. However, in our experiments, we find that the adaptation converges across the more than 100 test volumetric images across different anatomies and TDs, and across multiple runs of the experiments.
Fig.~\ref{fig_convergence_results} shows the convergence behaviour of the test-time adaptation. The top row in the figure demonstrates that the adaptation convergences reliably for different TDs. The bottom row shows the correlation between (a) the Dice between the predicted and ground truth segmentation and (b) the Dice between the DAE input and DAE outputs. It can been seen that these two Dice scores are correlated, thus justifying the choice of using the latter for determining the optimal I2NI parameters.}

\section{Discussion}\label{sec:discussion}
\noindent We proposed a method for cross-scanner and cross-protocol robustness in medical image segmentation by building on the ideas of test time CNN adaptation~\cite{wang2018interactive} and using denoising autoencoders to increase plausibility of predicted segmentations~\cite{larrazabal2019anatomical}. Our experiments show that the proposed method can yield promising improvements while segmenting images from completely unseen scanners and / or protocols. In this section, we elaborate on some avenues that could be potentially interesting for further research and for ultimately closing the gap to the benchmark - i.e. training a separate CNN for each scanner / protocol.

\vspace{0.1cm} \noindent \textbf{Noising strategy used for DAE training}: One of the main assumptions of our work is that the incorrect segmentations predicted by a CNN on an unseen TD can be considered to be from the training input distribution of the DAE, and therefore, that the output of the DAE is reliable. In this work, we chose a heuristic strategy for corrupting segmentation labels that the DAE seeks to denoise. We believe that the DAE performance can be improved if a better strategy can be devised to obtain noisy labels from the trained segmentation CNN. A potential way of doing this might be to train the segmentation CNN on the SD without data augmentation and then to use the predictions of this CNN on intensity transformed SD images (for instance, via gamma transformations) as noisy segmentations.

\vspace{0.1cm} \noindent \textbf{Assumption that DAE outputs are samples from the posterior $P(Z | Z^n)$}: Although this is common in the DAE literature~\cite{bengio2013generalized}, we claim that this is an assumption because given a noisy segmentation, the DAE is trained to output only one clean segmentation rather than several possible denoised segmentations. An interesting question is whether training the DAE to get such behaviour~\cite{baumgartner2019phiseg} might benefit the segmentation performance on unseen TDs.
    
\vspace{0.1cm} \noindent \newtextRtwo{\textbf{Prior on normalized images}: We trained segCNN, $S_{\theta}(N_{\phi}(.))$, with extensive data augmentation (DA) and then adapted $N_{\phi}(.)$ for each test image. It must be noted that even for a SD image, the output of $N_{\phi}(.)$ is an intermediate representation that is free to be different from the SD image itself. Therefore, we do not expect the test-time adaptation to tune $N_{\phi}(.)$ such that it acts as a translator from the TD to the SD. In future work, additional constraints may be imposed on this intermediate normalized representation, to further guide the test-time adaptation.}
    
\vspace{0.1cm} \noindent \textbf{Affine registration for large domain shifts}: The proposed atlas-based initialization for large domain shifts was only evaluated with brain images. Additionally, as both the SD (T1w) and TD (T2w) images used from the HCP dataset were already rigidly aligned, the affine registration step for atlas creation could be skipped. Such affine registration would be required for other TDs of brain images, as well as while applying the method to large domain shifts in other anatomies. We believe that not requiring deformable registration but only alignment with linear transformations might facilitate applications in other anatomies as well, but we leave this evaluation to future work.

\vspace{0.1cm} \noindent \textbf{Test-time adaptation in PGM framework}: A possible extension of our method might be to consider test-time adaptation of a supervised CNN in the probabilistic graphical model (PGM) framework used in unsupervised learning methods~\cite{van1999automated}. This would entail the use of an explicit prior model $P(Z)$ as well as a likelihood model $P(X|Z)$, with the test-time adaptation being driven with an aim of maximizing the resulting posterior $P(Z | X)$.

\vspace{0.1cm} \noindent \textbf{Time required for test-time adaptation}: The per-image flexibility offered by our method comes at the cost of the additional time required for such adaptation. After the first image of a particular scanner / protocol, the adaptation requires about 12 minutes for each 3D image, with our experimental implementation. Despite potential for improvement in terms of time efficiency, the proposed test-time adaptation does introduce an additional optimization routine for each test image and thereby compromises on the fast-inference advantage of CNNs. Nonetheless, we believe that such a time requirement is relatively modest and reasonable for general usage in clinical practice.
    
\section{Conclusion}\label{sec:conclusion}
\noindent In this work, we proposed a method for domain generalization for medical image segmentation in the context of domain shifts pertaining to scanner and protocol changes.
The method consists of two main ideas.
Firstly, we introduce an adaptable per-image normalization module into a segmentation CNN. We believe that such per-image adaptability may be crucial for developing robust analysis tools that can be deployed in the clinic.
Secondly, the proposed test time adaptation is driven by using denoising autoencoders, that incentivize plausible segmentation predictions.
Experiments with multiple datasets and anatomies demonstrate the promise and generality of the method, over other approaches such as data augmentation, meta-learning and unsupervised domain adaptation.

\section*{Acknowledgments}
This work was supported by: 1. The Swiss Platform for Advanced Scientific Computing (PASC), that is coordinated by the Swiss National Supercomputing Centre (CSCS), 2. Clinical Research Priority Program Grant on Artificial Intelligence in Oncological Imaging Network from University of Zurich, and 3. Personalized Health and Related Technologies (PHRT), project number 222, ETH domain. We also thank NVIDIA corporation for their GPU donation.

\bibliographystyle{IEEEbib}
\bibliography{ref}

\end{document}